\documentclass[useAMS,usenatbib]{mn2e}

\usepackage{graphicx}

\newcommand{\aap}{A\&A}
\newcommand{\mnras}{MNRAS}
\newcommand{\apj}{ApJ}
\newcommand{\apjl}{ApJL}
\newcommand{\apjs}{ApJS}
\newcommand{\araa}{ARA\&A}

\catcode`\@=11
\def\gsim{\ifmmode{\mathrel{\mathpalette\@versim>}}
    \else{$\mathrel{\mathpalette\@versim>}$}\fi}
\def\lsim{\ifmmode{\mathrel{\mathpalette\@versim<}}
    \else{$\mathrel{\mathpalette\@versim<}$}\fi}
\def\@versim#1#2{\lower 2.9truept \vbox{\baselineskip 0pt \lineskip 
    0.5truept \ialign{$\m@th#1\hfil##\hfil$\crcr#2\crcr\sim\crcr}}}
\catcode`\@=12
\def\msun{\hbox{$M_\odot$}}
\def\y1{\hbox{${\rm yr}^{-1}$}}

\title[A Multiwavelength Consensus on the Main Sequence of Star-Forming Galaxies at $\lowercase{z}\sim 2$]{A Multiwavelength Consensus on the Main Sequence of Star-Forming Galaxies at $\lowercase{z}\sim 2$} 

\author[G. Rodighiero, A. Renzini, E. Daddi et al.]
{\parbox{\textwidth}{\raggedright G. Rodighiero$^{(1)}$\thanks{E-mail: \texttt{giulia.rodighiero@unipd.it}},
A. Renzini$^{(2)}$,
E. Daddi$^{(3)}$,
I. Baronchelli$^{(1)}$,
S. Berta$^{(4)}$,
G. Cresci$^{(5)}$,
A. Franceschini$^{(1)}$,
C. Gruppioni$^{(6)}$,
D. Lutz$^{(4)}$,
C. Mancini$^{(2)}$,
P. Santini$^{(7)}$,
G. Zamorani$^{(6)}$,
J. Silverman $^{(8)}$,
D. Kashino$^{(9)}$,
P. Andreani$^{(10)}$,
A. Cimatti$^{(11)}$,
H. Dom\'inguez S\'anchez$^{(12)}$,
E. Le Floch$^{(3)}$,
B. Magnelli$^{(4,13)}$,
P. Popesso$^{(4)}$,
F. Pozzi$^{(11)}$}\vspace{0.4cm}\\
\parbox{\textwidth}{\raggedright $^{(1)}$Dipartimento di Fisica e Astronomia, Universit\`a di Padova, vicolo dell'Osservatorio 3, I--35122 Padova, Italy.\\
$^{(2)}$INAF - Osservatorio Astronomico di Padova, vicolo dell'Osservatorio 5, I-35122 Padova, Italy.\\
$^{(3)}$CEA-Saclay, Service d'Astrophysique, F-91191 Gif-sur-Yvette, France.\\
$^{(4)}$Max-Planck-Institut f\"{u}r Extraterrestrische Physik (MPE), Postfach 1312, D-85741 Garching, Germany. \\
$^{(5)}$INAF - Osservatorio Astronomico di Arcetri, largo E. Fermi 5,  I-50127 Firenze, Italy.\\
$^{(6)}$INAF - Osservatorio Astronomico di Bologna, via Ranzani 1, I-40127 Bologna, Italy.\\
$^{(7)}$INAF - Osservatorio Astronomico di Roma, via di Frascati 33, I-00040 Monte Porzio Catone, Italy.\\
$^{(8)}$Kavli Institute for the Physics and Mathematics of the Universe (WPI), Todai Institutes for Advanced Study, The University of Tokyo, Kashiwanoha, Kashiwa 277-8583, Japan. \\
$^{(9)}$Division of Particle and Astrophysical Science, Graduate School of Science, Nagoya University, Nagoya 464-8602, Japan.\\ 
$^{(10)}$ESO, Karl-Schwarzschild-Strasse 2, D-85748 Garching, Germany.\\
$^{(11)}$University of Bologna, Department of Physics and Astronomy (DIFA), V.le Berti Pichat, 6/2 - 40127, Bologna, Italy.\\
$^{(12)}$Departamento de Astrof\'{\i}sica, Facultad de CC.  and F\'{\i}sicas, Universidad Complutense de Madrid, E-28040, Madrid, Spain.\\
$^{(13)}$Argelander-Institut f\"{u}r Astronomie, University of Bonn, auf dem H\"{a}gel 71, D-53121 Bonn, Germany.}}

\begin{document}

\date{}

\pagerange{\pageref{firstpage}--\pageref{lastpage}} \pubyear{2012}

\maketitle

\label{firstpage}
\begin{abstract}
{We compare various star formation rate
(SFR) indicators for star-forming
galaxies at $1.4<z<2.5$ in the COSMOS field. The main focus is on the
SFRs  from the far-IR (PACS-{\it Herschel} data)
with those  from the ultraviolet, for galaxies selected
according to the BzK criterion.  FIR-selected samples
lead to a vastly different slope of the SFR-stellar mass ($M_*$)
relation, compared to that
of the dominant {\it main sequence} population as measured from the
UV, since the FIR selection
picks predominantly only a minority of {\it outliers}.
However,  there is overall agreement between the main sequences
derived with the two SFR
indicators,  when stacking on the PACS maps the BzK-selected
galaxies.   The resulting logarithmic slope of the
SFR-{$M_*$} relation is $\sim0.8-0.9$, in agreement with that derived
from the dust-corrected UV-luminosity.
Exploiting deeper 24$\mu$m-{\it Spitzer} data we have characterized a
sub-sample of  galaxies  with
reddening and SFRs  poorly constrained, as they are very faint in the $B$ band.
The combination of Herschel with Spitzer data have allowed us to largely break
the age/reddening degeneracy for these intriguing sources, by distinguishing whether
a galaxy is very red in B-z because of being heavily dust reddened, or whether
because star formation has been (or is being)  quenched.
Finally, we have compared our SFR(UV) to the SFRs derived by stacking
the radio data and to those derived from the
H$\alpha$ luminosity of a sample of star-forming galaxies at
$1.4<z<1.7$.  The two sets
of SFRs are broadly consistent  as they are with the SFRs derived from
the UV and by stacking the corresponding PACS data in various mass bins.}

\end{abstract}


\section{Introduction}
\label{intro}
Most galaxies at high redshifts are very actively forming stars, with
star formation rates (SFR) of order of hundreds $M_{\odot} $ yr$^{-1}$ being
quite common. In the local Universe, instead, galaxies with such high
SFRs are very rare and are called ``ultraluminous infrared galaxies'' (ULIRG, with $L_{\rm
 IR}>10^{12}\,L_\odot$, \citealt{Sanders88}). Such objects are caught in a transient, {\it
  starburst} event, likely driven by a merger having boosted both their SFR and their far-IR luminosity.
By analogy,  also such high-redshift galaxies were first regarded as starburst objects, until
it became apparent that data were suggesting a radically different picture.

A first suspicion that a new paradigm was needed came from the
discovery that over $80\%$ of a ``$BzK$'' $K$-band selected sample of $z\sim
2$ galaxies were actually qualifying as ULIRGs (\citealt{Daddi05}). Clearly, it was very unlikely that the vast majority of
galaxies had all been caught in the middle of a transient event. As later shown, at high
redshifts sustained SFRs ought to be the norm rather than the exception.

This was indeed demonstrated in a series of seminal papers (\citealt{Elbaz07}; \citealt{Daddi07}, \citealt{Noeske07}), showing the
existence of a tight correlation between SFR and stellar mass $M_*$,
with
\begin{equation}
SFR\propto f(t)M_*^{1+\beta},
\end{equation}
which is followed by the majority of star-forming (SF) galaxies, with
a dispersion of $\sim 0.3$ dex, both at high redshifts (references
above) and in the local Universe (\citealt{Brinchmann04}). Thus,
following Noeske et al. (2007) the correlation is called the {\it Main
  Sequence} (MS) of SF galaxies. Here $f(t)$ is a declining function
of cosmic time (an increasing function of redshift).
Furthermore, no signs of mergers have been
found through dynamical measurements in many high redshift
star forming galaxies (e.g. \citealt{Forster09}, \citealt{Cresci09}, \citealt{Law09}).
Implying that most SF galaxies are in a quasi-steady SF
regime, the existence of the MS has several important ramifications. It
dictates a very rapid stellar mass growth of galaxies at early times,
paralleled by a secular growth of their SFR itself (e.g., \citealt{Renzini09}; \citealt{Peng10}), quite at odds with the widespread assumption of
exponentially decling SFRs (as argued by e.g., \citealt{Maraston10}
and \citealt{Reddy12}). Even more importantly, the slope
$\beta$ controls the {\it relative} growth of high mass vs. low-mass
galaxies, thus directly impinging on the evolution of the galaxy stellar
mass function (\citealt{Peng13}, see also \citealt{Lilly13}).

While the existence of the MS is generally undisputed, its
slope and width may differ significantly from
one observational study to another, depending on the sample selection
and the adopted SFR and stellar mass diagnostics. Selecting galaxies
in a passband that is directly sensitive to the SFR (such as the
rest-frame UV or the far IR) automatically induces a Malmquist bias in
favor of low-mass galaxies with above average SFRs, thus flattening
the resulting SFR$-M_*$ relation. This effect is clearly seen in
Herschel FIR-selected samples, where formally $\beta 
\simeq -1$, but where only a tiny fraction of galaxies are detected at low
stellar masses, i.e., those few really starbursting ones (\citealt{Rodighiero10a, Rodighiero11}).
This Malmquist bias  has also been recognized in simulations (\citealt{Reddy12}).

If redshifts are measured spectroscopically, the final sample may still suffer a
similar bias  even if the original photometric selection ensured  a
mass-limited input catalog. Indeed, at low masses the success
rate of getting redshifts may be higher if the SFR is
above average, and it may be lower at high masses if such galaxies are
heavily extincted. Again, both these effects will tend to flatten the
SFR$-M_*$ relation.

For example, \citet{Reddy06} and \citet{Erb06}
found no positive correlation at all between SFR and stellar mass (i.e., $\beta
\sim -1$) for a spectroscopic sample of UV-selected galaxies at $z\sim 2$, whereas
\citet{Reddy12} found an almost perfectly linear relation ($\beta
\sim 0$)  for a sample of similarly selected galaxies, when taking into account the result of their simulation.

On the other hand, other biases may tend to steepen the derived
SFR$-M_*$ relation.  Indeed, the mere selection  of SF galaxies
(e.g., by color or by a SFR cut) may preferentially exclude massive galaxies
with below-average SFR.

At low redshifts the most suitable SFR indicator is the H$\alpha$
luminosity (see \citealt{Dominguez12}), which is available for the extremely large sample of SDSS
galaxies for which  \citet{Brinchmann04} and \citet{Peng10} got
$\beta\simeq -0.1$ using this SFR indicator. However,  already at
relatively low redshift H$\alpha$ moves out of the optical range
and \citet{Noeske07}  resorted on  the 24 $\mu$m flux together
  with the less reliable [OII] luminosity as SFR
indicators, deriving $\beta\simeq -0.3$ for their sample of $0.2<z<0.7$
galaxies. Conversely, \citet{Elbaz07} got $\beta\simeq -0.1$ for
star-forming galaxies at $z\sim 1$ using the Mid-IR (24 $\mu$m flux) as a SFR indicator.
The same $\beta\simeq -0.1$ slope was then found by \citet{Daddi07} for a
mass-selected sample of $z\sim 2$ galaxies, using the
extinction-corrected UV luminosity to measure SFRs.
Finally, by combining 24 $\mu$m detection and SED fitting, \citet{Santini09} found a similar value of
$\beta\simeq -0.15 $ for star-forming galaxies at $z\simeq2$.

Stacking 1.4 GHz radio data in various mass bins proved to be another effective way of
measuring the slope (and normalization) of the SFR$-M_*$ relation,
with \citet{Pannella09} getting $\beta\simeq 0$ for galaxies at
$z\sim 2$. However, stacking the same radio data \citet{Karim11}
found $\beta\simeq -0.4$ for their sample of SF galaxies,
having defined them as those bluer than $(NUV-r^+)_{\rm rest}=3.5$, a
definition that following Ilbert et al. (2010) includes both
``active'' ($(NUV-r^+)_{\rm rest}<1.2$)   and ``intermediate'' ($1.2<(NUV-r^+)_{\rm rest}<3.5$)
SF galaxies. Restricting to ``active'' galaxies, Karim et al. found
$\beta$ oscillating between $\sim 0$ and $\sim -0.2$ with no obvious trend with redshift.

In summary, these examples illustrate that the derived value of the
slope $\beta$ critically depends on several assumptions and adopted
procedures, namely:

\par\noindent$\bullet$
The starting photometric selection. For example magnitude/flux
limited, (multi-)color selection or mass limited.  

\par\noindent$\bullet$
The procedure to measure redshifts. Spectroscopic redshifts add to the
photometric selection their instrument-specific selection function (i.e., the
success rate as a function of photometric magnitudes and
colors). Photometric redshifts are less biasing in this respect,
modulo their occasional  catastrophic failure.

\par\noindent$\bullet$
The criterion to separate SF from non-SF galaxies. As mass quenching
dominates at high redshifts (\citealt{Peng10}), a SF criterion that may
retain galaxies on their way to be quenched would bias $\beta$ towards
more negative values.

\par\noindent$\bullet$
The adopted SFR indicator, including in it the procedure to estimate
the dust extinction, if required.

\par\noindent$\bullet$
The explored mass range, as the slope at low masses might differ from
that at high masses.

In this paper we derive the SFR$-M_*$ relation for a mass-complete
sample of SF galaxies at $1.4<z<2.5$ using a variety of SFR
indicators, such as the UV continuum, the H$\alpha$ luminosity, the
Mid-IR 24 $\mu$m flux, the FIR luminosity, and the radio luminosity,
then stacking data when appropriate to derive the average SFR$-M_*$
relation for the mass-limited sample.

Throughout the paper we use a \citet{Salpeter55} initial mass function (IMF) and we assume
$H_0=70$ km s$^{-1}$, $\Omega_{\Lambda}=0.75$, $\Omega_{M}=0.25$ and AB magnitudes.

\section{Observations and sample selection}

Homogeneous samples of sources would ideally be required to
compare the results of different SFR estimators in a  meaningful way.  Unfortunately, this
is normally quite difficult as the selection functions tend to bias
samples from various surveys having different depths, spectral ranges
and selection wavelength (see e.g., \citealt{Wuyts11}).  In this paper
we combine far-IR-selected (i.e., SFR-selected) and near-IR-selected
(as a proxy to $M_*$-selected) star-forming samples in the COSMOS field
(\citealt{Scoville07}), having both UV- and IR-based SFR
determinations (both mid- and far-IR).  A fraction of them have been spectroscopically
observed to measure the H$\alpha$ emission line luminosity, providing
an additional indicator of SFR.  Radio observations from the literature
are used to extend the comparison of widely used SFR tracers.  We
first describe the datasets used, the sample selections and the SFR
and $M_*$ measurements.

We consider only galaxies within the redshift range of $1.4\lsim
z\lsim 2.5$, based either on spectroscopic or photometric redshifts.

\subsection{Herschel far-IR samples}

We start from the sample of PACS/{\it Herschel} observations in the COSMOS field described by \citet{Rodighiero11},
over 2.04 square degrees and down to a $5\sigma$ detection, above confusion limits of 8 and 17~mJy at 100 and 160 $\mu$m,
respectively  (\citealt{Lutz11}). Photometry was carried out by PSF-fitting at 24$\mu$m prior positions.  
The detection limits correspond to $\sim100 \, M_\odot$ yr$^{-1}$, $\sim
200\,M_\odot$~yr$^{-1}$ and $\sim 300\, M_\odot$ yr$^{-1}$, respectively
at $z=1.5$, 2 and 2.5.
Over a common area of 1.73 square degrees we cross-matched the PACS
detections with the IRAC-selected catalog of \citet{Ilbert10}, so to obtain UV-to-8$\mu$m photometry, accurate photometric redshifts and
stellar masses by SED fits as described in \citet{Rodighiero10b}. 
At $z\sim2$ the  sample of  \citet{Ilbert10} is complete in mass above
$\sim10^{10}M_{\odot}$  for star-forming galaxies (see their Table
3). FIR $8-1,000\,\mu$m luminosities ($L_{\rm IR}$) 
are derived from PACS fluxes using a set of empirical templates  as described in \citet{Rodighiero10b} and \citet{Rodighiero11}. 
In this work IR luminosities are always converted to SFR as
SFR$[M_{\odot}{\rm yr}^{-1}]=1.7\times10^{-10}L_{\rm IR}[L_{\odot}]$ (\citealt{Kennicutt98},
hereafter SFR(FIR)).
By adopting different templates or codes, consistent SFR estimates are obtained with no bias and a scatter of $\sim 0.15$~dex  
(that represents the typical error associated to our SFRs, see also \citealt{Berta13}).  
The dataset includes in total 576 PACS-detected galaxies with
$1.4<z_{\rm phot}<2.5$.

\subsection{BzK samples}
\label{BzK}
We use the $K$-band selected sample of $1.4<z<2.5$ star-forming galaxies down to $K_{\rm s,AB}<23$ in the
COSMOS field (\citealt{McCracken10}) selected according to the  
criterion (\citealt{Daddi04}) designed to pick star-forming galaxies
at these redshifts (the so-called star-forming BzK, or sBzK), i.e.,
those sources with:
\begin{equation}
(z-K)_{\rm   AB}-(B- z)_{\rm AB}\ge -0.2.
\end{equation}

The passively evolving BzK (or pBzK) are not discussed in this paper,
apart from the possible contamination of the formal sBzK sample.
Stellar masses  have been computed following the same procedure as in
\citet{Daddi04} and \citet{Daddi07}, adopting the empirically calibrated relation based on the BzK
photometry alone:
\begin{equation}
{\rm log}(M_*) = -0.4(K_{\rm tot}-19.51)+0.218(z-K)-0.499.
\end{equation}
In spite of its simplicity, the procedure gives stellar masses which
with a 0.3 dex scatter are in excellent agreement with those
obtained with full fledged SED fits.
For all selected sBzK the SFRs are estimated from the UV rest-frame luminosity corrected for
dust extinction  (hereafter SFR(UV)), with reddening being inferred
from the slope of the UV continuum as in \citet{Daddi07}.
UV-based SFRs reach down to few $\msun$~yr$^{-1}$ at $z\sim2$. 
The final sBzK sample includes a total of 25,574 sources in the redshift range $1.4<z<2.5$.
For the rest of the paper we will use consistently these stellar
  masses unless stated otherwise. In \citet{Rodighiero11}  we verified that they are
fully consistent with those used in \citet{Rodighiero10b} and in the
previous subsection. 

\par\noindent$\bullet$ {\it good-}sBzK: with a formal error $\delta$log[SFR(UV)]$<0.3$ dex (21,375 sources);
\par\noindent$\bullet$ {\it bad-}sBzK:  with a formal error $\delta$log[SFR(UV)]$>0.3$ dex (4,199 sources).

The relative uncertainty on SFR(UV) is formally derived by propagating the errors on the optical photometry of each source,
in particular from the $B$ and $z$ bands used to compute  $E(B-V)$ and then to derive a dust-corrected SFR(UV) (see \citealt{Daddi04}):
\begin{equation}
\delta{\rm log}[SFR(UV)]=\sqrt{E1^2 + E2^2 + E3^2 +E4^2},
\end{equation}
with $E1=0.6\times \delta B$, $E2= \delta z$, $E3=0.1$ and
$E4=(0.75\times0.06\times(1+z_{\rm phot}))$,
where $\delta B$ and $\delta z$ are the photometric errors on the $B$
and $z$ magnitudes and 0.1 is a term that  accounts for the 
error on the estimate of the total magnitude of the galaxy. Note that
the different coefficients of $\delta B$ and $\delta z$ stem from the
$B$ magnitude entering twice in the calculation of the SFR, once to
estimate the reddening and once to measure  the observed UV
luminosity, whereas the $z$ magnitude affects only the reddening estimate.
The last  term ($E4$) accounts for the uncertainty of the photometric
redshift where we assume the typical $\delta z_{\rm phot}/(1+z_{\rm phot})\sim0.06$
and 0.75 is an empirical coefficient depending on the typical UV slope and luminosity distance of the objects.

The $good$-sBzK sample selection represents $\sim$84\% of the whole sBzK population, and can be considered as a 
criterium to select reliable SFR(UV) estimates (at least at the limits of the COSMOS survey). 
 
We should mention that among the $\sim$16\% of the  {\it bad-}sBzK sources, $\sim$5\% of them are undetected in the $B$ band
(at the COSMOS survey depth), implying that their SFR(UV) can not simply be computed or classified.
These sources will be considered in the {\it Herschel} stacking analysis and will be still included in the {\it bad-}sBzk  classification.
However, when showing the SFR(UV) for the single sources in the {\it bad-} and {\it good-}sBzK sample, they will not be reported.

In principle, the BzK criterion may introduce a bias by selecting only part of the star-forming galaxies in the explored redshift range. The same would also do a pure photometric redshift selection, given the sizable number of photometric redshifts which are grossly discrepant with respect to spectroscopic redshifts (especially at $1.4<z<1.8$, cf. \citealt{Ilbert10}). For this reason we have inspected the Ilbert et al. catalog,
finding that  $\sim6\%$ of $1.4<z_{\rm phot}<2.5$, $M_*>10^{10}$ M$_\odot$  objects are missed by the BzK criterion (including both star-forming and passive sources). However, a major fraction of them lie very close to the line defined by Equation (2), hence just small photometric errors have driven them out of the sBzK domain. The others are likely cases in which $z_{\rm phot}$ fails catastrophically.  Thus, we believe that a sample of sBzK-selected galaxies with $1.4<z_{\rm phot}<2.5$ is more robust than either a purely sBzK- or a purely $<z_{\rm phot}$-selected sample. In any case, slope and dispersion of the SFR(UV)$-M_*$ main sequence  are not appreciably affected by the inclusion of this minority population.

\subsection{H$\alpha$ spectroscopic sample}
\label{Ha}
As part of a Subaru telescope survey with FMOS (Fiber Multi-Object
Spectrograph) in its high-resolution mode ($R\sim2600$), the  sBzK
population in the inner deg$^2$ of the COSMOS field has been targeted
to detect $H{\alpha}$ in emission from galaxies at $1.4\lsim z\lsim 1.7$
(\citealt{Kashino13,zahid13}, Silverman et al. in prep.). Sources have been
selected from the sample described in Section \ref{BzK} to have
stellar masses $>10^{10}\msun$ and to belong to the {\it good}-sBzK population.

The measured  H$\alpha$ luminosities for the 162 best quality
  (flag=2) detections are converted to SFR  (hereafter
SFR(H$\alpha$)) with the Kennicutt (1998) relation,
SFR(H$\alpha$)$[M_{\odot}{\rm yr}^{-1}]=3.03\times 10^{-8}L({\rm H}\alpha)/L_\odot$.
The H$\alpha$ luminosity has been corrected for extinction applying the average
$A_{H\alpha}-M_*$  relation from
the Balmer decrement of FMOS spectra stacked  in mass bins (Kashino et al. 2013).

\section{SFR from various  indicators} 
\label{sfrs}
In this section we present a systematic comparison of SFRs from
various widely used SFR indicators, focusing in particular on their
effect on the SFR-stellar mass relation of our program galaxies at
$1.4<z<2.5$.

\begin{figure*}
\centering
\includegraphics[width=18cm]{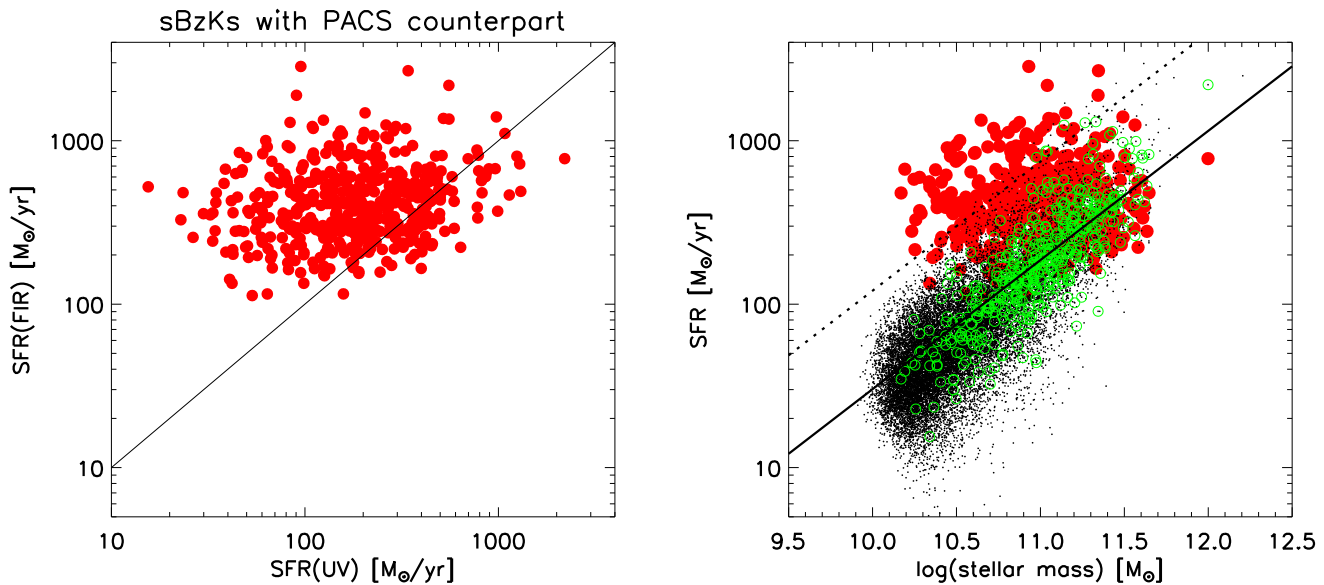}
\caption{{\it Left panel}: comparison of SFR(UV) and SFR(FIR) for a
  sample of 473 sBzK at $1.4<z<2.5$ with a PACS/{\it Herschel} detection in the COSMOS field.
The {\it right panel} shows the SFR-stellar mass relation for various
samples, namely:  parent sBzK sample (the
so-called {\it good} subsample, see text for details, small black
dots) for which SFR(UV) is reported, 
the PACS-detected sBzK sources shown in the upper panel, with
SFR(FIR), and for the same group of galaxies the green
open circles represent the corresponding
SFR(UV). The solid (dotted) line indicates the MS  
(SFR(UV)$=4\times$SFR(MS)) relation at $z\sim2$ (\citealt{Rodighiero11}).
}
\label{SFR_UV_FIR}
\end{figure*}

\subsection{Far-Infrared versus Ultraviolet SFRs}
\label{sfr_ir}
Figure \ref{SFR_UV_FIR}  (left panel)  compares the SFRs from the
far-IR and from the ultraviolet, i.e., SFR(FIR) vs.  SFR(UV).
We used the sample of 473 sBzK at  $1.4<z<2.5$ in the COSMOS field for which a PACS counterpart is available.
It is apparent that the calorimetric indicator, able to almost completely reveal the hidden SFR, provides systematically higher
values than SFR(UV), in particular at SFR(UV)$\lsim 300\; M_{\odot}$
yr$^{-1}$\footnote{
This can be the case if the detected rest frame UV is emitted from a relatively unobscured region of the galaxy, whereas most of the SF activity 
is heavily extincted.}. 
This is commonly interpreted as an underestimate of  dust extinction as derived
from the UV slope (i.e., from the $B-z$ color, as in \citealt{Daddi07},
having potentially an important impact on the slope and scatter of the star-forming main sequence.
This is shown in the right panel of Figure \ref{SFR_UV_FIR}, where
we show the mass-SFR relation for the parent sample of {\it good}-sBzK
(small black dots). To emphasize the effect of different SFR
indicators in shaping the MS, for the PACS sources shown  in 
the left panel,  the right panel displays both their SFR(FIR) (red filled circles) and their SFR(UV)
(green open circles), while using  the same stellar mass.
By relying only on SFR(FIR), one
gets a flat SFR$-M_*$  relation, with $\beta\simeq -1$ in Equation
(1). Such a flat  relation is the direct result of having
selected galaxies using a far-IR flux limited sample, which translates
indeed into a SFR-limited sample.  On the other hand, the UV indicator provides a much steeper relation
(solid line in Figure \ref{SFR_UV_FIR}, with $\beta=-0.21$,
{\it good}-sBzK only, \citealt{Rodighiero11}).
This  illustrates the point made in the Introduction, about how
different the slope of the MS  can result when using different
selection criteria  or SFR indicators.

\begin{figure*}
\centering
\includegraphics{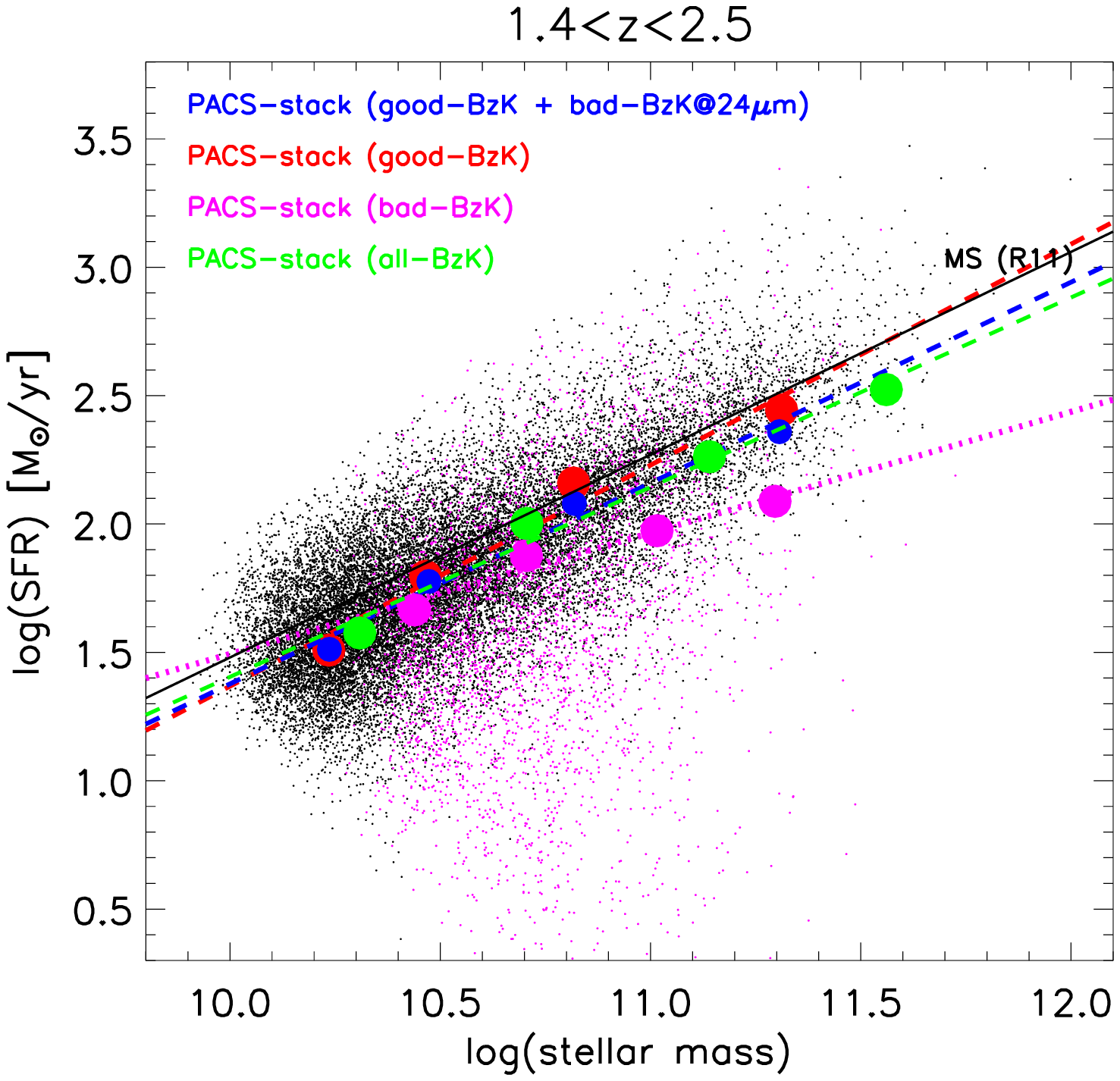}
\caption{The SFR-stellar mass relation for star-forming galaxies at $1.4<z<2.5$ is shown for
  various samples: the small black dots represent the parent {\it good}-sBzK.
  Most of these sources have a reliable estimate of
  extinction from the $(B-z)$ color, and thus a
  reliable SFR from the UV.  The complementary sample of sBzK for which
  SFR(UV) is much less reliable ({\it bad}-sBzK), is shown with small magenta dots.  These sBzK samples
  have been split into four mass bins. The red filled circles show the
  average SFR derived by stacking on the PACS maps the {\it good}-sBzK
  (with the corresponding best linear fit shown as a dashed red line),
  while the magenta circles refer to the stacking results for the {\it bad}-sBzK (with the corresponding best linear fit shown as a
  dashed magenta line).  Green filled circles represent the SFR obtained by
  stacking the whole sBzK population in the four different mass bins (with the corresponding best linear fit shown as a
  dashed green line). The blue circles correspond to the stack of
    the {\it good}-sBzK sample plus the {\it bad}-sBzK which are
    detected at 24 $\mu$m, and the corresponding best fit is shown as
    the blue dashed line.
For each mass bin the error bars on SFR are derived from the bootstrap statistical stacking analysis and are smaller than the symbol sizes.
 The solid black line represents the best fit to the Main Sequence derived by \citet{Rodighiero11}.}
\label{PACS_stack}
\end{figure*}

This apparent  discrepancy derives from the vastly different
number of galaxies recovered by the two selections, the
Herschel/SFR-selected sample and the sBzK/mass-selected sample. As made
clear in Figure \ref{SFR_UV_FIR}  (right panel), for  log($M_*)\lsim
11$ only a few sBzK galaxies are
individually detected by Herschel, and include (part of) the $\sim 2\%$ {\it outliers} from the
MS as shown by \citet{Rodighiero11}. We interpreted these objects
as obscured starbursts, possibly driven by merging 
events or major disk instabilities, characterized by  high specific-SFR 
(sSFR=SFR$/M_*$), and where $E(B-V)$ and the SFR from the UV are
systematically underestimated.

On the other hand, the Herschel-COSMOS data at these redshifts do not reach
below SFR $\sim200 M{_\odot}$ yr$^{-1}$ and therefore to recover a
far-IR MS we must resort on
stacking the Herschel data at the location of sBzK-selected galaxies,
which represent a mass-selected sample. To this end,
we split the sBzK sample into four mass bins, and stack all PACS-undetected sBzK if  a residual
160$\mu$m map created by removing all PACS 160$\mu$m detections with
SNR$>3$ (stacking at 100$\mu$m does nwellot change our results). 
The stacking is performed using the IAS stacking library (\citealt{Bethermin10}), PSF-fitting photometry,
and applying an appropriate flux correction for faint, non-masked sources to the PACS stacks (\citealt{Popesso12}).
With this procedure, we derived the average  flux for each mass bin. Using the formalism introduced by \citet{Magnelli09}, 
that accounts both for detections and no-detections,  we then
converted these stacked fluxes into bolometric luminosities $L_{\rm IR}$
by adopting an average $K$-correction (\citealt{Chary01}) and then
into SFR through the  standard law of \citet{Kennicutt98}.
bog
The results of this procedure are presented in Figure \ref{PACS_stack}.
We considered the whole sBzK sample  at $1.4<z<2.5$, and then separately
the {\it good}- and {\it bad}-sBzK sub-samples,
represented by small black dots and small magenta dots, respectively.
The big red filled circles show the average SFR derived by stacking on
the PACS maps only the {\it good}-sBzK (with the corresponding best
linear fit shown as a dashed red line, slope
$\alpha=1-\beta=0.86\pm0.08$). The magenta circles refer instead to the stacking
results for the {\it bad}-sBzK (slope $\alpha=1-\beta=0.47\pm0.12$) whereas
the green data points represent the SFR obtained by stacking the whole
sBzK population, with the corresponding best linear fit shown as a
dashed green line (slope $\alpha=1-\beta=0.74\pm0.11$). 

Overall, there is a nice agreement of SFR(UV) and stacked SFR(FIR) for
the {\it good}-sBzK sample, largely amending  the discrepant results
when using only the individually PACS-detected sources
(Figure \ref{SFR_UV_FIR}).  The MS slope using SFR(FIR)
($\alpha=1-\beta=0.86\pm0.11$) is consistent within the errors
with that derived using SFR(UV)  ($\alpha$=$1-\beta$=0.79$\pm0.10$, \citealt{Rodighiero11}).
This argues for  the correlation of SFR(UV) and
SFR(FIR) to be fairly good for the general MS population at $z\sim2$, a
correlation that instead clearly  fails catastrophically 
for the most obscured starburst sources, which represent only few
percent of the star-forming galaxies at the same cosmic 
epoch  (Figure \ref{SFR_UV_FIR}).   Still, it is somewhat intriguing that
  for these galaxies (the green open circles in
  Figure  \ref{SFR_UV_FIR}) the `wrong' SFR(UV)  places them within the
  main sequence,  probably because the optical colours refer only the
  small fraction of the SFR which is not fully buried  in dust. 

\begin{figure}
\centering
\includegraphics[width=9cm]{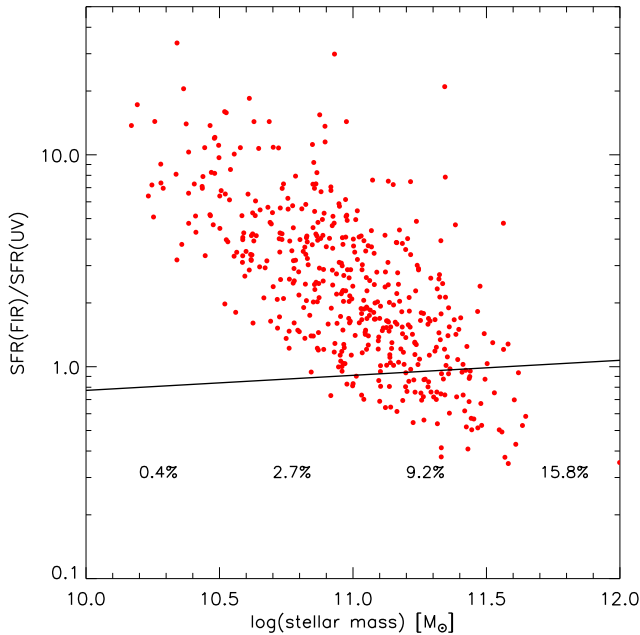}
\caption{The SFR(FIR)/SFR(UV) ratio for galaxies that are individually
  detected by Herschel/PACS over the COSMOS field (red points). The
  fractions of such detected sources over the parent {\it good}-sBzK population
  are given for four mass bins, each 0.5 dex wide. The  nearly
  horizontal line represents the ratio of the best
fit SFR(FIR)$-M_*$  and SFR(UV)$-M_*$  relations from Figure
\ref{PACS_stack}.
}
\label{gianni}
\end{figure}

Figure \ref{gianni} further illustrates and quantifies these
findings. The data points represent the SFR(FIR)/SFR(UV) ratio for the 
{\it good}-sBzK galaxies which are individually detected by the
Herschel/PACS PEP survey over the COSMOS field. At low masses this  ratio is very high ($\sim
10$) and decreases with increasing mass reaching near unity towards
the high mass end. However, at low masses only 0.4\% of the {\it
  good}-sBzK galaxies are detected in the infrared, i.e., only the
extreme outliers.
Then the fraction of FIR-detected galaxies increases with stellar
mass, reaching $\sim 16\%$ at the top end. This is still far from $100\%$, as the PEP data are not deep enough to recover all
galaxies even at the top mass end. Notice that the minimum measured
SFR(FIR) ($\simeq 200\,\msun\y1$) refers to $z=2$, and increases with
redshifts, whereas the completeness of the PEP catalog starts dropping
at substantially higher values (\citealt{Rodighiero11}). In deeper PEP
observations,
such as those on the GOODS-South field, the fraction of massive
galaxies which are detected does indeed approach 100\% (\citealt{Rodighiero11}).
A further confirmation that SFR(UV) does not systematically deviate
from SFR(FIR) comes from the stacking of the Herschel/PACS data
discussed above and illustrated in Figure \ref{PACS_stack}. The almost
horizontal line in Figure \ref{gianni} shows the ratio of the best
fit SFR(FIR)$-M_*$  and SFR(UV)$-M_*$ relations from Figure
\ref{PACS_stack}, thus emphasizing that both methods of deriving the
SFR are fully consistent for the vast majority of the galaxies, with
the exception of a lesser minority of outliers.

When including all sBzK in the far-IR comparison (green circles and
green line), the slope of the {\it Herschel} derived MS  ($\alpha=1-\beta=0.74\pm0.08$) is still largely overlapping
with that derived from the UV.
For what concerns the {\it bad}-sBzK sample alone, Figure
\ref{PACS_stack} indicates that at low masses ($M_*<10^{11}M_{\odot}$)
the mean SFR(FIR) is consistent with that of the most reliable SFR(UV)
sample, while at higher masses it is systematically lower, hinting for
a contamination by passive sources into the star-forming 
color selection. To check for this possibility in the next section we
consider the MIPS 24$\mu$m properties of these galaxies and we further
expand on this issue.

\subsection{Mid-Infrared versus Ultraviolet SFRs}
\label{24-UV}
A natural extension of the {\it Herschel} based SFR analysis includes
the widely used 24 $\mu$m MIPS/{\it Spitzer} flux density, that allows
one to reach lower SFRs than {\it Herschel}, although with the large
extrapolation required to estimate the total IR luminosity (e.g. \citealt{Elbaz07},
\citealt{Elbaz11}, \citealt{Wuyts11}). Since the earlier {\it
  Herschel} investigations it was realized that the 24 $\mu$m SFR
indicator was working very well up to redshift $\sim1$, while it
starts to fails at higher redshifts by overestimating somewhat the
true $L_{\rm IR}$ (\citealt{Nordon10}, \citealt{Nordon12}, \citealt{Rodighiero10b}, \citealt{Elbaz11}).
This is particularly critical at $z\sim2$, where the
PAH features enter the observed 24 $\mu$m pass-band.  More recently,
\citet{Magdis12} have undertaken a systematic study of the typical
SED of normal star-forming and starburst galaxies at $z\sim2$,
including both PACS and SPIRE/{\it Herschel} data in their
analysis. They found that the mean SED does not evolve along the MS at
$z\sim2$, while it differs for the starburst population (characterized
by a warmer dust component). Similar results are found also by \citet{Elbaz11}.  
These new investigations revamped the use of the 24
$\mu$m SFR indicator, ideally allowing the adoption of a universal SED
to extrapolate $L_{\rm IR}$ for MS sources.  Other recipes and methods
have been presented to recalibrate the 24 $\mu$m flux density (\citealt{Nordon12}; 
\citealt{Wuyts11}; see also \citealt{Berta13} for a summary).  
In this Section we adopt the MS templates of Magdis et al. (2012)
to extrapolate $L_{\rm IR}$ from the 24 $\mu$m flux densities.

\begin{figure*}
\centering
\includegraphics[width=15cm]{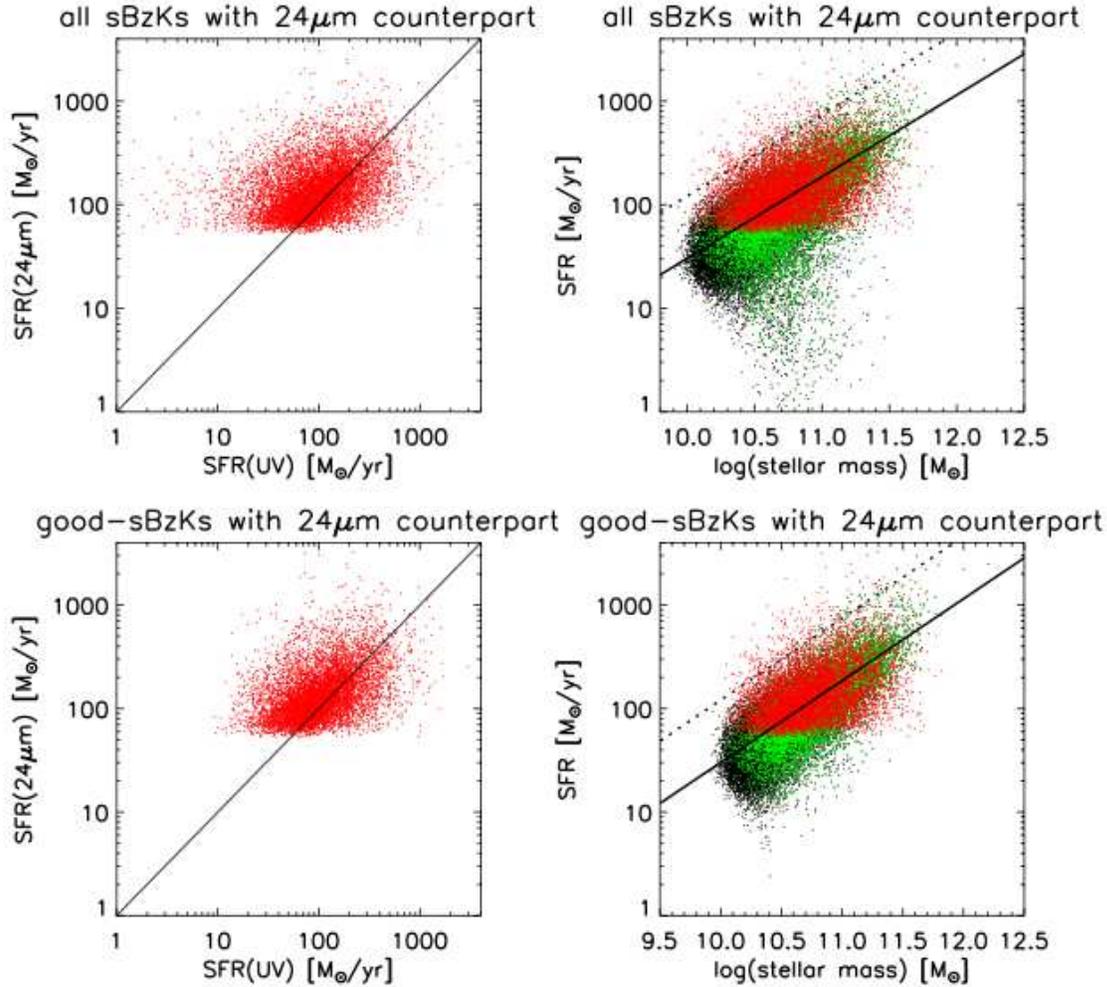}
\caption{{\it Left panels}: comparison of SFR(UV) and SFR(24$\mu$m)
  for the sample of sBzK  at $1.4<z<2.5$ with a MIPS/{\it Spitzer} 24$\mu$m detection (S$_{24\mu m} >60\mu$Jy) in COSMOS.
{\it Right panels}:  the SFR-stellar mass relation of star forming galaxies as
shaped by different SFR indicators: red points represent the
sBzK/MIPS sources shown in the left panels. The green points are
the same sources plotted with the corresponding SFR(UV). For
completeness, we  show also SFR(UV) for the parent sBzK sample (small
black dots). The solid (dotted) line indicates the best fit to the
Main Sequence as in Figure \ref{SFR_UV_FIR}.
{\it Upper panels} include all sBzK-selected galaxies, while {\it lower
  panels} report only the {\it good}-sBzK, for which a reliable estimate of the extinction is available from the UV slope.
}
\label{SFR_UV_24}
\end{figure*}

Following the same approach of Section \ref{sfr_ir}, in Figure
\ref{SFR_UV_24} (left panels) we compare SFR(UV) with SFR($24_{\mu \rm m}$) for the
sample of sBzK in COSMOS with a 24 ${\mu {\rm m}}$ counterpart
brighter than S$_{24\mu \rm m} >60\mu$Jy.
The corresponding differences induced in the MS relation are instead
shown in the right-hand panels. We separate the analysis including all sBzK
(top panels) and only the {\it good}-sBzK (bottom panels).  Red  points
represent the sBzK/MIPS-detected sources shown in the left
panels, with SFR from $L_{\rm IR}$ extrapolated from the 24$\mu$m flux
density. The green points are the same sources plotted with the
corresponding SFR(UV). For completeness, we report also SFR(UV) for the parent sBzK
sample (small black dots). The solid (dotted) line indicates the MS
($\times4$MS) relation at $z\sim2$ (Rodighiero et al. 2011), as in
Figure \ref{SFR_UV_FIR}.  The considered flux limit allows us to reach
SFR as low as $\sim$60$M_{\odot}$ yr$^{-1}$, diving well into the MS,
but it still shows the  almost flat SFR$-M_*$ relation  which is
typical of SFR-selected samples (see Figure \ref{SFR_UV_24}, right panels).

This SFR(UV)-SFR($24_{\mu \rm m}$) relation including all sBzK sources
is rather  dispersed, showing, as for the SFR(UV)-SFR(IR), an excess of
objects with SFR($24_{\mu{\rm m}})>$ SFR(UV), particularly for
SFR(UV)$<100\ M_\odot$ yr$^{-1}$. Indeed, the penalty of a wrong
(underestimated) extinction correction is evident for the sBzK sources
with  a less reliable SFR(UV) (the {\it bad}-sBzK):
in the top-left panel the tail at low SFR(UV) ($\lsim 10\, M_{\odot}$
yr$^{-1}$) is populated by these objects, that instead largely disappear
when considering only the {\it good}-sBzK (bottom-left panel).
In this case  the MS based on SFR($24_{\mu {\rm m}}$) nicely overlaps with
the UV-based one, with the advantage of unraveling also the starburst
sources (with SFR$>4\times$SFR(MS))  that remain unidentified when using SFR(UV).
Thus, the mid-IR reveals this population of main sequence {\it
  outliers}, as does the far-IR (\citealt{Rodighiero11}), but the extrapolation required to derive $L_{\rm IR}$ from the
24 $\mu$m  flux density still makes the far-IR information a more
direct and effective mean to estimate the global SFR of high-redshift dusty sources.

The {\it bad}-sBzK which are detected at 24 $\mu$m are clearly
  star forming and therefore should be considered together with the
  {\it good}-sBzK when stacking the
  {\it Hershel} data to derive the slope and zero point of the main
  sequence. The result is illustrated in Figure \ref{PACS_stack} (blue
  circles and dashed line) and the corresponding slope is 
  $\alpha=1-\beta=0.80\pm 0.07$. We consider this as our best possible
  estimate of the main sequence slope at $z\sim 2$.

\begin{figure*}
\centering
\includegraphics[width=15cm]{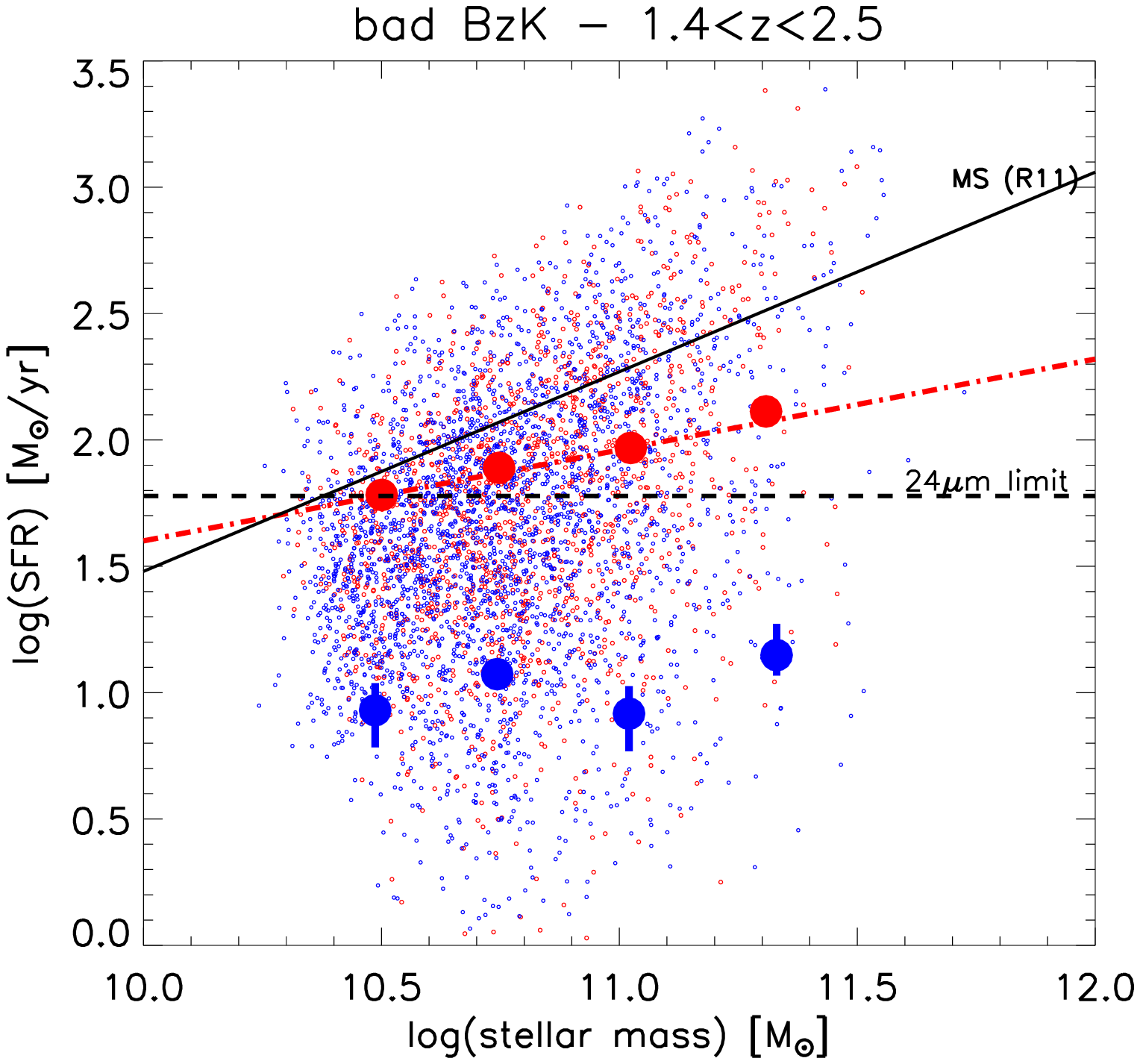}
\caption{The SFR(UV)-stellar mass relation for the {\it bad}-sBzK with red
  symbols referring to the MIPS 24 $\mu$m detected sources and the
  blue symbols to the 24 $\mu$m undetected ones. The corresponding large
circles show SFR(FIR) having stacked the {\it Herschel}/PACS data in
four mass bins. The dashed horizontal line corresponds to the MIPS 24
$\mu$m sensitivity limit over the COSMOS field at $z=2$ and the solid line is
the same as in Figure 1. 
For each mass bin, error bars on SFR are derived from the bootstrap statistical stacking analysis 
and presented with the same color coding (if bigger than the symbol sizes).
}
\label{stack__PACS_badEBV_no24}
\end{figure*}

The 24 $\mu$m flux density allows us also to better characterize the
population of the {\it bad}-sBzK. For example, among the 3219
  sBzK galaxies with $M_* > 10^{11}M_\odot$ in our sample there are
  787 
  such objects, $\sim 60\%$ of which (467) 
  are not detected at 24 $\mu$m, corresponding to a SFR upper
limit of $\sim 60\, M_{\odot}$ yr$^{-1}$.  This is well below the SFR
of massive MS galaxies and we infer that most of the 467 {\it
  bad}-sBzK are likely to be well on their way to be quenched. This is
further reinforced by the result of stacking the Herschel 160 $\mu$m
data, separately for the 24 $\mu$m detected and undetected {\it
  bad}-sBzK, as displayed in Figure
\ref{stack__PACS_badEBV_no24}. Clearly, on average the 24 $\mu$m
undetected {\it bad}-sBzK galaxies lie well below the MS, whereas the
24 $\mu$m detected ones lie appreciably below the MS and exhibit a
shallower slope ($\alpha=1-\beta=0.36\pm 0.04$).  We recall that our
sBzK selection is supposed to pick star-forming galaxies, whereas now
we have evidence that out of the original 25,574 sBzK $\sim 4199$ of
them ($\sim 16\%$) are likely to be quenched or on the way to be
quenched.  Of course, we cannot exclude that some of these
  galaxies are experiencing a temporary downward excursion from the
  main sequence and will return to it in the future, i.e.,
  representing a tail of the main sequence itself. Data cannot
  distinguish between such objects and truly quenching ones. 
 However, we note from Figure 6 that the bad-sBzK
are confined to relatively high masses, where
galaxies are faint in the B-band because they are either
heavily reddened or because they are quenched or on the way to be quenched.
In the former case they should be detected at 24 $\mu$m but they are not, which suggests they are actually
quenched. Note also the absence of low mass
bad-sBzK, while there should be many of them if
they would represent a tail of the main sequence
distribution. Moreover, Figure \ref{stack__PACS_badEBV_no24} shows that when stacking the FIR data for the 24 $\mu$m-undetected
sources their average SFR is well below the main sequence values (from $\sim 5$ to $\sim 30$ times
below) which suggests that the vast majority of them are likely to be quenched or on their way to be quenched.
   We believe this illustrates the capability of this
multiwavelength approach of singling out MS galaxies as well as the
starburst and quenched outliers on either side of the MS. {\it In
  summary, the {\it bad}-sBzKs include a mixture of actively
  star-forming galaxies and others which may be fully quenched or with
  SFRs well below the MS, though the distinclion between these two
  latter subclasses would need deeper data.}

\subsection{BzK sources selected for being star-forming actually not being so}
\begin{figure*}
\centering
\includegraphics[width=9cm,height=12cm]{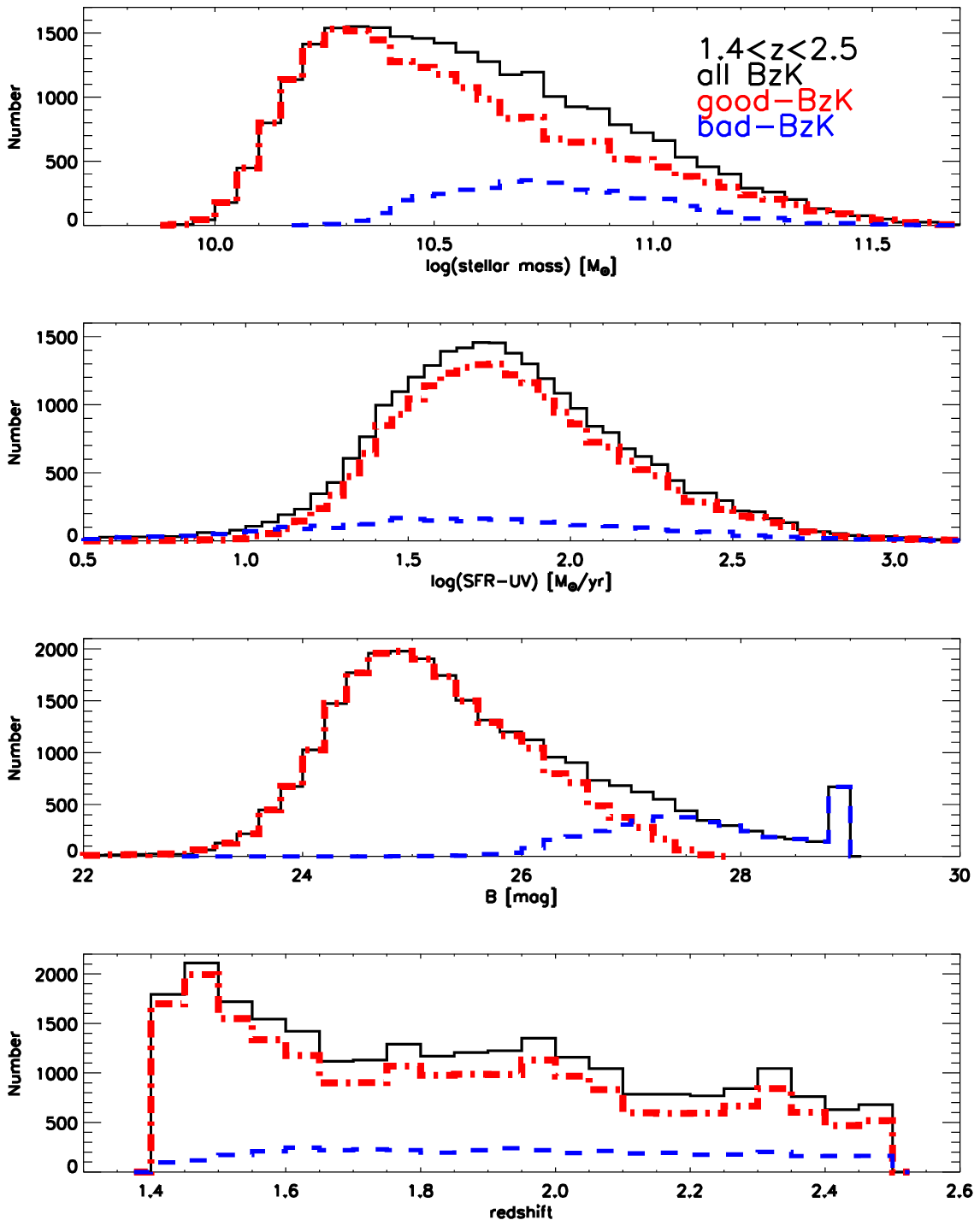}
\caption{Statistics of the sBzK sample at $1.4<z<2.5$ in the COSMOS field, as a function of stellar mass ({\it top panel}), SFR(UV) 
({\it second panel from top}), observed $B$ magnitude ({\it second panel from top}) and redshift distributions ({\it bottom panel}). We
report the distribution of sBzK sources with reliable SFR(UV)
(i.e. {\it good}-sBzK, dot-dashed red lines), the {\it bad}-sBzK
(dashed blue lines), and the total distribution (solid black lines).
Objects undetected in the $B$ band are all assigned to the faintest
bin of the $B$-band histogram. The redshift distribution of the
  three populations is shown in the bottom panel.
}
\label{bad-good}
\end{figure*}

In our previous analysis we made an intensive use of the sBzK
classification based on the relative error on SFR(UV) to understand the
quality and limits of the SFR derived solely from the rest-frame
UV. We have seen that, formally, when considering only reliable
sources (i.e., $\sim$84\% of the sBzK COSMOS sample, those with $\delta$log([SFR(UV)])$<0.3$ dex) then
SFR(UV) is in very good agreement with SFR(IR) for the vast majority
of the galaxies.  To better characterize
the properties of these various sBzK classes, we present in Figure
\ref{bad-good} the distribution of their stellar masses (top panel),
SFR(UV) (second panel from top), $B$ magnitudes (third panel from top) and redshifts (bottom panel). We report
separately the distributions for the {\it good}-sBzK (dot-dashed red
lines),  the {\it bad}-sBzK (dashed blue lines), and the total distribution
(solid black lines).  Notice that the mass distribution starts
dropping at $\sim 2\times 10^{10}\,\msun$, which we consider the
completeness limit of our sample. This is nearly twice as large as the mass
limit  of the  \citet{Ilbert10} $1.5<z<2$ sample, as our sample
extends to $z=2.5$. 

As expected, the intrinsic larger errors on
SFR(UV) (as propagated from formal errors on the original photometry)
is mostly related to the faintness of these sources in the $B$ band\footnote{
The faintness of the bad-sBzK in the B band does not primarily
derive from the relative distance of such class, since their
redshift distribution is almost flat over the whole range
(see Figure  \ref{bad-good}, bottom panel), although the ratio of bad- to good-sBzKs
moderately increases with redshift.}:
the peak of the observed $B$-band distribution is $\sim$2.5 mag
brighter for the {\it good}-sBzK.  On the contrary, the SFR(UV)
distributions for the two samples span the same range, with the {\it
  bad} sources presenting only a tiny fraction excess at low SFR(UV),
as already revealed in Figure \ref{SFR_UV_24}.  However, this low-SFR
tail does not impact on the main trend for SFR(UV)-SFR(IR), as
revealed by the PACS stacking analysis (Figure \ref{PACS_stack} and
Section \ref{sfr_ir}), and it consists of a mixture of two opposite
kinds of sources: 1) passive sources that appear to fulfill the
Equation 2 star-forming (sBzK)  selection because of their large error
in  the $B$-band magnitude, and 2) very
obscured/starburst objects for which SFR from the UV catastrophically
fails (as it does for a small minority of the {\it good}-sBzK as well).

In this respect, we can notice on Figure
\ref{stack__PACS_badEBV_no24} that quite many of the most  massive {\it bad}-sBzK
exhibit a SFR(UV) well in excess of the MS values, whereas
their average SFR(IR) from stacking the Herschel data falls well below
the MS. We conclude that the population of the  {\it bad}-sBzK is
indeed a mixture of obscured starburst and of quenching galaxies, with
the former ones dominating at lower masses and the latter ones
dominating at high masses. This trend can be readily understood when
considering that the fraction of (starburst) MS outliers ($\sim 2\%$)
is fairly independent of stellar mass (\citealt{Rodighiero11}), hence
low mass outliers must be more numerous, whereas at high masses the 
{\it mass-quenching} mechanism of \citet{Peng10} must be proceeding
at full steam at these redshifts.  We also notice that for the {\it  bad}, 24 $\mu$m 
undetected sBzK the procedure to get the SFR from
UV is delivering a SFR about an order of magnitude too high because it
mistakes the red $B-z$ color as due to reddening, while it is due to
old age. Thus, the {\it bad} fraction of the star-forming  selection is
effectively contaminated by a  number of galaxies which are
either already quenched or being quenched. 
These amount to $\sim 60\%$ of the {\it bad}-sBzK sample of galaxies more
massive than $10^{11}\,M_\odot$, or $\sim 15\%$ of the whole sBzK
sample above this mass limit. Ironically, for most  {\it bad}-sBzK,
many of those with very high SFR(UV) are actually quenched (the small
blue point in Figure \ref{stack__PACS_badEBV_no24} with SFR(UV)$>>
60\, M_\odot$ yr$^{-1}$) and many of those with very
low SFR(UV) are actually starbursting (the small red points in the
same figure with SFR(UV)$<<60\, M_\odot$ yr$^{-1}$).
In this regard, it is worth emphasizing that the {\it bad}-sBzK which
are actually quenched were clearly misclassified as star forming in
the first place. At the faintest $B$ magnitudes the error $\delta B$
can be so large  to qualify a galaxy as a sBzK according to Equation (2),
while the real $B$ magnitude would have actually classified it as a 
passively evolving, pBzK galaxy. Finally, we notice that the {\it
  flattening} of the main sequence towards high masses, especially
when including the {\it bad}-sBzK, is likely due to a large fraction
of the most massive galaxies being already on their way to be quenched
(e.g., \citealt{Whitaker12}; C. Mancini et al., in preparation).

\begin{figure*}
\centering
\includegraphics{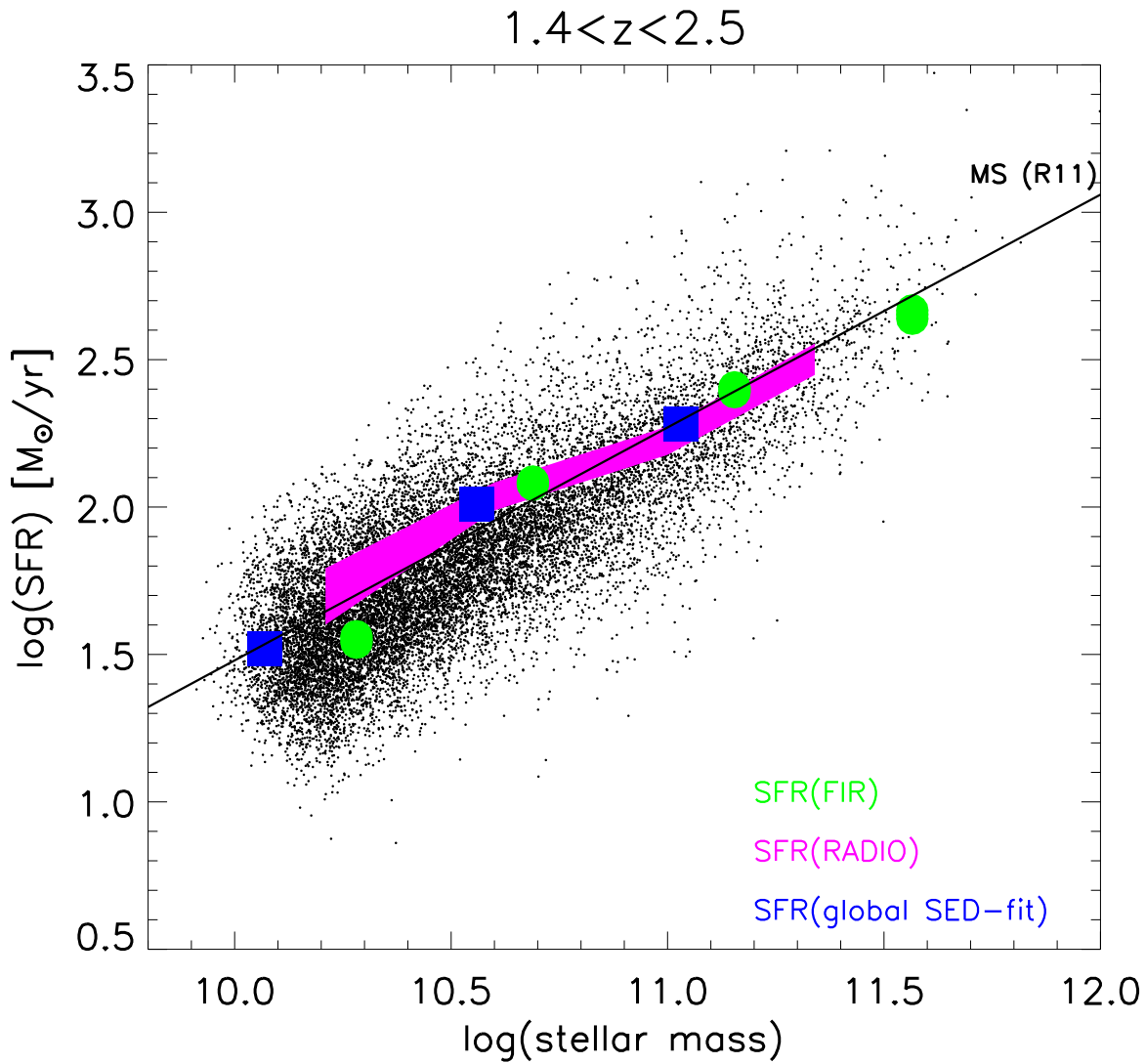}
\caption{
Comparison in the SFR-stellar mass 
plane  of the SFR from stacked radio data (magenta
shaded region, Karim et al. 2011) and stacked far-IR data (green
data points, as in Figure 2 for the {\it good}-sBzK). 
We also report
SFR(global SED-fit) for  sBzK sources as derived in
three mass bins using the  median SEDs (from near-IR up to
submillimeter) as derived by \citet{Magdis12}. The small black
points refer to the SFR(UV) for the {\it good}-sBzK.
}
\label{SFR_radio}
\end{figure*}

\subsection{Radio and global near IR-to-submillimeter SED fitting}
\citet{Pannella09} and \citet{Karim11} have measured the
average SFR in various mass and redshift bins by  stacking the COSMOS 1.4 GHz
radio continuum emission, by using either $BzK$ or IRAC mass-selected
samples, respectively.

In Figure \ref{SFR_radio} we directly compare the results of 
Karim et al. (2011) with ours in the common
redshift interval ($1.4<z<2.5$).  The figure shows the SFR(UV) for sBzK-selected sources (small
black points) and the stacked SFR(IR) from PACS (green filled circles)  while the magenta shaded
region corresponds to the radio analysis by \citet{Karim11}.  To
convert the average 1.4 GHz luminosities into average SFRs
Karim et al.  used the calibration of the radio-FIR correlation by
\citet{Bell03}. We have rescaled their data to the IMF adopted in this
paper.  The slope and normalization of the stacked radio SFRs are in
good agreement with both the PACS ones and the UV based.  This
result is not surprising, given the well known tight correlation
between the radio and far-IR luminosities. 

An indirect approach that combines various ingredients consists in
integrating the median SED of sBzK in various mass bins along the MS.
As anticipated in Section \ref{24-UV}, \citet{Magdis12} have
obtained average mid- to far-IR SEDs of $z\sim2.0 $ MS galaxies in three
stellar mass bins, derived by stacking observed data from 16$\mu$m up
to 1100$\mu$m. They also provide the total IR luminosities of each
template for each mass bin, that we converted into an average SFR with
\citet{Kennicutt98}.  The results of this exercise  are shown as
blue filled squares in Figure \ref{SFR_radio}, and the resulting
SFR-mass relation  is fully consistent
with the MS defined by UV, Herschel and radio data,
providing a further support to the concordance of average SFR
indicators at $z\sim2$. It is certainly reassuring
that by applying different criteria for mass-selected samples and
different SFR indicators  we obtain consistent results in such a
wide range of stellar masses.

\begin{figure*}
\centering
\includegraphics{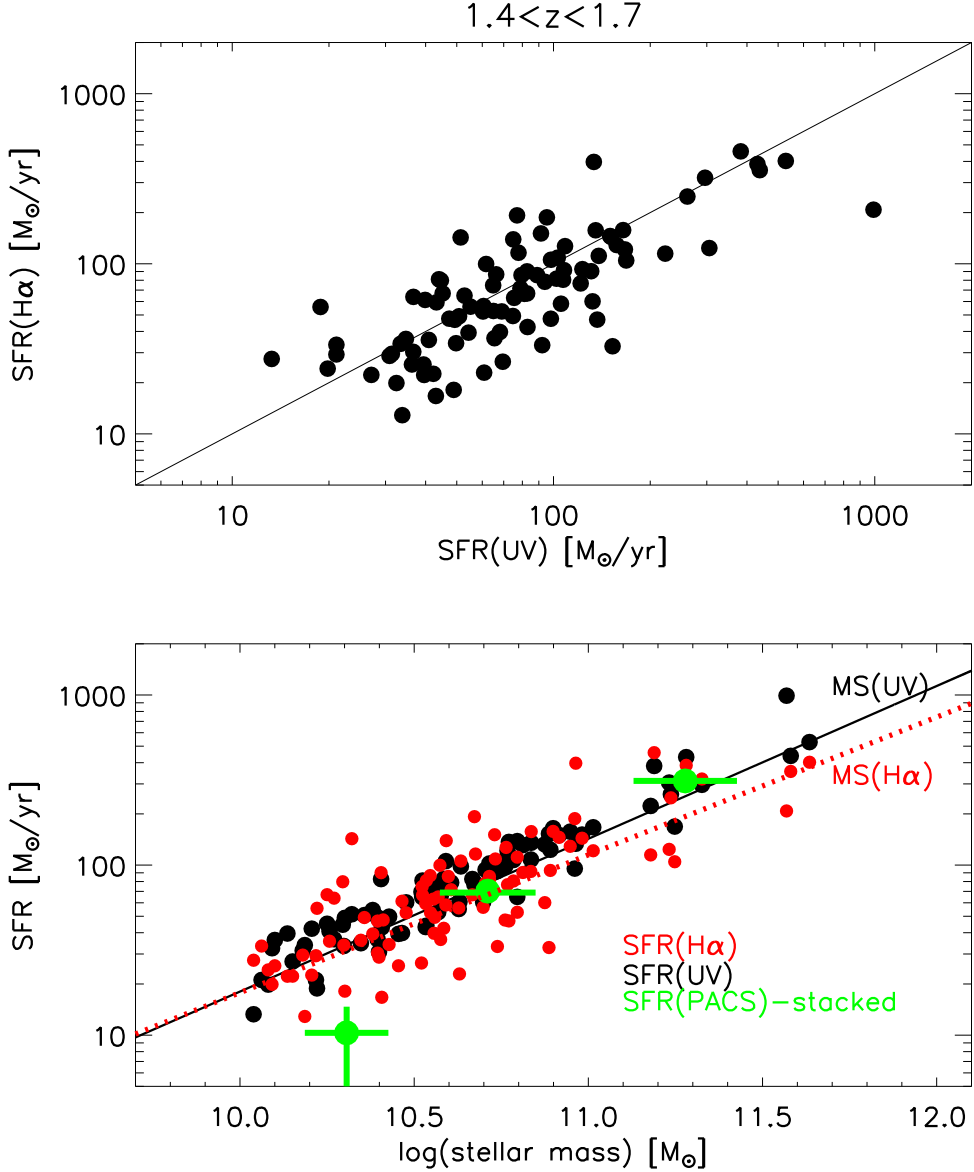}
\caption{{\it Top panel:} Comparison of SFR(UV) and SFR(H$\alpha$) for a sample of sBzK sources at $1.4<z<1.7$ 
spectroscopically observed  with FMOS/{\it Subaru}
and for which a direct measure of the H$\alpha$
luminosity is available (\citealt{Kashino13}). Extinction corrections
are derived from the average 
$A_{H\alpha}$-stellar mass linear relation derived by Kashino et al. (2013).
{\it Bottom panel:} For the same sources,  the stellar SFR-stellar
mass relation is shown.
For each source we show the SFR(UV) (black circles) and the corresponding SFR(H$\alpha$) (red circles).  
By stacking these sources in  three mass bins on the PACS maps, we
obtained a mean value of the corresponding SFR(IR)  (plotted as green symbols). 
The width of the stacked data along the $x$-axis represents the standard deviation of the mass distribution in each bin.  
The typical uncertainties on SFR are derived by the bootstrap stacking procedure.
The solid black line is the best-fit relation obtained by linear interpolation to the sBzK population with their SFR(UV) at $1.4<z<1.7$, 
while the red dotted line is the best-fit relation in the same redshift interval obtained by \citet{Kashino13} from SFR(H$\alpha$).
}
\label{SFR_Ha_UV}
\end{figure*}

\subsection{SFR from H$\alpha$ Luminosity}
As mentioned in Section \ref{Ha}, a fraction of the star-forming 
with photometric redshifts in the range $1.4\lsim z\lsim 1.7$ have
been selected as targets for the Intensive Program at the Subaru
telescope with the FMOS near-IR spectrograph (J. Silverman et al. in
preparation; \citealt{Kashino13}). The first observing runs in the $H$-long
band have provided the detection of H$\alpha$ and spectroscopic
redshifts for 271 galaxies, 168 of them having high quality (flag = 2)
line detections.  Kashino et al. include in their analysis
also FMOS spectroscopy in the $J$-band, to assess the
level of dust extinction by measuring the Balmer decrement using
co-added spectra. They found that the extinction at H$\alpha$
is an increasing function of stellar mass and they
provide a linear empirical relation between these two quantities, 
as $A_{H\alpha} \simeq 0.60+1.15\, ({\rm log}[M_*/\msun] -10)$.  In this work we adopt this recipe to compute
dust-corrected SFR(H$\alpha$) (see Section \ref{Ha} for details), and we
limit our analysis to the 168 flag=2 sources.  We first compare the
derived SFR(H$\alpha$) and SFR(UV) in Figure \ref{SFR_Ha_UV} (top
panel), showing a good correlation between the two independent SFR
measures. The median SFR(UV) for this sample is 
$\sim 20\%$ higher than SFR(H$\alpha$), suggesting that  the Balmer
decrement may underestimate  the total extinction  (see
discussion in \citealt{Kashino13}).  To better understand this trend,
we have stacked on the 160 $\mu$m PACS maps these sources in three mass bins. This
is presented in the usual mass-SFR plot in the bottom panel of
Figure \ref{SFR_Ha_UV}, showing for each source SFR(UV)
(black circles) and the corresponding SFR(H$\alpha$) (red circles),
while the green symbols show the SFR(IR) from the stacked PACS data.  The width of
the stacked bins along the $x$-axis represents the standard deviation of
the mass distribution in each bin.  
The uncertainties on the  stacked SFR are derived from the bootstrap stacking procedure, and
in the two higher mass bins they are of the size of the green data points. The stacked
SFR(IR) in the smaller mass bin is lower then the corresponding average SFR(UV) and SFR(H$\alpha$) 
but we believe this is not significantly so. Contrary to the two more massive bins no individual sources are detected in the FIR and therefore
the bootstrap stacking procedure underestimates the error bars.
The solid black line is the MS relation obtained by linear
interpolation to the sBzK population with their SFR(UV) at $1.4<z<1.7$
(slope $\alpha=0.90\pm0.11$), while the red dotted line is the best-fit relation in the
same redshift interval obtained by \citet{Kashino13} from
SFR(H$\alpha$) (slope $\alpha=0.81\pm0.04$).  The UV indicator is more consistent with
SFR(IR) than the H$\alpha$ luminosity, in particular at higher masses,
where the flatter relation derived by SFR(H$\alpha$) might suggest
that the extinction correction derived from the Balmer decrement is
more uncertain for massive objects (cfr. \citealt{Kashino13}).  A
slight bias is also present in the H$\alpha$ sample as at low masses
objects with above average SFR(UV) were selected for the FMOS
observations. A more
comprehensive investigation of dust extinction affecting the intrinsic
luminosity of emission lines will be presented at the completion of the whole FMOS survey.

\section{Discussion and Conclusions}
We have used the COSMOS multiwavelength database to derive masses and
star formation rates of $1.4<z<2.5$ galaxies using a variety of SFR indicators, such as the UV
luminosity, the far-IR ($8-1,000\;\mu$m) luminosity, and the 24 $\mu$m
flux. For galaxies in the redshift range $1.4<z<1.7$ we have also
estimated the SFR using the H$\alpha$ line luminosity. Stellar masses
have been derived from  SED fits using UV-to-8$\mu$m photometry. The
same set of masses have been used irrespective of the SFR indicator,
so to isolate the effect of using different indicators. Of course, the
characterization of high-redshift galaxies may also be biased by the
specific procedure to measure stellar masses, but exploring this
aspect is beyond the scope of the present paper, that is instead
focused on the effects of using different SFR indicators, specifically
on the slope of the SFR$-M_*$ relation followed by the majority of
galaxies and known as the Main Sequence of star forming galaxies.

We have shown that the selection criteria to pick star-forming
galaxies have a profound effect on the
slope of the SFR-$M_*$ relation. Using observables that are directly linked to the SFR
(such as the mid- and the far-IR)
the resulting SFR$-M_*$ relation tends to be essentially flat, but one
recovers only a small fraction of the galaxies selected to produce a
mass-limited sample. We show in particular that for $M_* \lsim
10^{11}\,M_\odot$ the 160 $\mu$m 
selection (from {\it Herschel})  picks predominantly galaxies for
which the SFR derived from the UV luminosity falls largely short of
that indicated by their far-IR luminosity. Arguably, in
such extreme cases this is due to the inability of the slope of
the rest-frame  UV continuum to estimate the true dust extinction
affecting the bulk of the star formation in such galaxies. Such a selection
picks predominantly starbursting outliers from the MS, but fails to pick
the vast majority of star-forming galaxies in the same mass range, whose
far-IR luminosities are below the {\it Herschel } detection limit.

To take advantage of the positive aspects represented by the
reliability of far-IR based SFRs on one side, and of mass-limited
samples on the other, we recour to stacking the {\it Herschel} data in
various mass bins, showing that the logarithmic slope of the SFR$-M_*$
relation derived from such stacks is in excellent agreement with that
derived from the dust-corrected UV luminosity, and is in the range
$\sim 0.8-0.9$.

The considerations on the SFRs derived from the far-IR luminosity apply
as well to the SFRs derived from the 24 $\mu$m flux, which actually in
COSMOS reaches to lower SFR levels. This offers the opportunity to
better characterize a sub-sample of star-forming sBzK-selected galaxies,
i.e., those for which reddening and SFRs are poorly constrained by the
observed rest-frame continuum, here nicknamed the {\it bad}-sBzK,
i.e., those very faint in the $B$ band. About 50\% of them are
detected at  24 $\mu$m and therefore qualify as star-forming galaxies.
Stacking their {\it Herschel}/PACS 160 $\mu$m data shows they
are close to the MS, though with a slightly flatter slope. However,
particularly interesting are the {\it bad}-sBzK which are {\it not}
detected at  24, 100 and 160 $\mu$m, and whose stacked PACS data show they
have SFRs well below the MS  (the blue points in Figure
  \ref{stack__PACS_badEBV_no24}) and therefore qualify for being 
quenched (or quenching) galaxies. Therefore, the combination of {\it
  Herschel} with  {\it Spitzer}  data have allowed us to break the
age/reddening degeneracy for sBzK-selected galaxies, thus
distinguishing whether a galaxy is very red because of being heavily
dust reddened, or whether it is very red because star formation has
been quenched.

Finally, we have compared our SFR(UV) to the SFRs  derived from the
H$\alpha$ luminosity of a sample of sBzK-selected galaxies at
$1.4<z<1.7$ observed with FMOS at the Subaru telescope.  The two sets
of SFRs are broadly consistent with each other as they are with the
SFRs derived by stacking the corresponding PACS data in two mass bins.
As a result, also the SFR$-M_*$ relation using SFR(H$\alpha$) values
is consistent with that derived from the other SFR indicators.

The reassuring conclusion is that a wide variety of SFR indicators,
such as the rest-frame UV continuum, the mid- and the far-IR, the 1.4
GHz radio flux and the
H$\alpha$ luminosity all give consistent results when applied to
samples as close as possible to be mass-selected samples. The slope of
the main sequence can vary between $\sim 0.8$ and $\sim 1$, depending
on the specific selection criterion and on the adopted SFR indicator,
which all must introduce a small bias. Perhaps the most intriguing of
such biases comes from how star-forming galaxies are identified as
such, as especially at high masses a non trivial fraction (almost
$\sim 15\%))$ of sBzK-selected galaxies (selected for being star forming) turns out to be already 
quenched or well on their way to be quenched, as indeed expected to
happen thanks to the {\it mass quenching} process, an effect that
tends to flatten the slope of the main sequence. Ironically, many 
 {\it bad}-sBzK with low SFR(UV) turn out to be very powerful mid-
 and far-IR sources and are starbursting MS outliers.

\section*{Acknowledgments}
GR, IB and AF acknowledge support from the University of Padova from ASI (Herschel Science Contract  I/005/07/0).         
AR acknowledges funding support from a INAF-PRIN-2010 grant.
ED acknowledges funding support from ERC-StG grant UPGAL 240039 and
ANR-08-JCJC-0008.
GC acknowledges support from grant PRIN-INAF 2011 "Black hole growth and AGN feedback through the cosmic timeÓ
AC acknowledges the MIUR PRIN 2010-2011 "The dark Universe and the
cosmic evolution of baryons: from current surveys to Euclid".
This work was supported by World Premier International Research Center
Initiative (WPI Initiative), MEXT, Japan.

PACS has been developed by a consortium of institutes led by MPE (Germany) and including UVIE 
(Austria); KU Leuven, CSL, IMEC (Belgium); CEA, LAM (France); MPIA (Germany); INAF- 
IFSI/OAA/OAP/OAT, LENS, SISSA (Italy); IAC (Spain). This development has been supported by the 
funding agencies BMVIT (Austria), ESA-PRODEX (Belgium), CEA/CNES (France), DLR (Germany), 
ASI/INAF (Italy), and CICYT/MCYT (Spain).

We thank the anonymous referee for a careful reading and valuable comments,
which have significantly contributed to improve the clarity of the paper.

\label{lastpage}

\end{document}